# Optimizing AI-Driven Messaging for Type 2 Diabetes Management: Insights from Patient Preference Elicitation

**Angela Mastrianni, PhD[1,2], Defne Levine, MPH[1], Katerina Andreadis, MA[1,2], Lynn Xu, MPH[1], Priscilla D'Antico, PhD, MPH[1,3], Antoinette Schoenthaler, EdD[1,3], Devin Mann, MD, MPH[1,2]**

**[1]Department of Population Health, NYU Grossman School of Medicine, New York, NY, USA; [2]Department of Health Informatics, NYU Langone Health, New York, NY, USA; [3]Institute for Excellence in Health Equity, NYU Langone Health, New York, NY, USA**

**Abstract**

*Generative AI (GenAI) allows for improved user experience within conversational agents for diabetes management by supporting dynamic, context-aware conversations. In this study, we elicited patient preferences for the communication style of a GenAI-based conversational agent (uMatter) developed to support diabetes management. We conducted an online survey with 125 individuals with type 2 diabetes. The survey included a discrete choice experiment to evaluate participant preferences for different types of messaging attributes. The survey also elicited participant perceptions and feedback on the messages from uMatter. We found significant preference heterogeneity for the inclusion of emojis within the messages. Additionally, qualitative findings indicated that participants had different desired personas and communication styles for the conversational agent. We propose strategies from recent human-computer interaction and natural language processing research that can be used to design GenAI-based conversational agents that align with the communication preferences of patients.*

**Introduction**

Type 2 diabetes is a significant and growing health problem. In addition to being a leading cause of death and disability in the United States, glycemic control among individuals with type 2 diabetes has continued to worsen in recent years, despite the availability of evidence-based guidelines for effective care.[1,2] One key barrier to optimal glycemic control is inadequate support for diabetes self-management outside of the primary care setting.[3] Nearly half of individuals diagnosed with diabetes report receiving no education on managing diabetes, with limited clinician and diabetes educator workforce capacity restricting the time and resources available to provide personalized education and support to patients.[4]

Mobile health technologies (mHealth) can be used to support diabetes self-management, but challenges with personalization and engagement have limited their success. Conversational agents can be incorporated within mHealth platforms to provide real-time support to patients through natural language user interfaces for self-monitoring, medication adherence, and lifestyle change.[5,6] Initial studies suggest that these agents are generally accepted by patients and can be effective in improving health outcomes (e.g., glycemic control).[5,7] However, the effectiveness of the agents is often constrained by limited personalization and interactivity, which reduce patient engagement.[5,8] Patients have expressed a desire for conversational agents that both generate personalized content and respect their highly-individualized communication preferences.[9]

Generative AI (GenAI) offers a promising approach to improve user experience within conversational agents for diabetes management.[10] Traditional conversational agents have often used rule-based approaches (e.g., decision trees, keyword matching) to manage conversation flow and generate responses, limiting flexibility and depth of personalization.[11] In contrast, GenAI models (e.g., GPT, Claude, Gemini) can be integrated into conversational agents to produce dynamic, context-aware, and more human-like content, enabling greater personalization and responsiveness to individual patient needs. Unlike rule-based systems that operate within predefined conversational pathways, GenAI-enabled agents can generate a wide range of communication styles and tones that may be tailored in real time to align with patient preferences, clinical context, and engagement patterns. However, this flexibility also introduces new design considerations. Without clear guidance, generated content may not consistently align with patient expectations or preferred communication styles. As a result, further research is needed to systematically elicit and quantify patient preferences to inform the design and prompting strategies for GenAI-enabled conversational agents supporting diabetes management.[12,13]

## Methods

We used a mixed-methods approach to elicit patient preferences for a GenAI-based conversational agent (uMatter), a component of the randomized controlled trial iMatter2, which aims to test an mHealth tool designed to support patients with self-management of their diabetes. An online survey was conducted with 125 individuals with type 2 diabetes. In the first section of the survey, participants completed discrete choice experiments (DCEs) designed to evaluate preferences for inclusion of three messaging attributes: (1) a signature, (2) emojis, and (3) explicit motivational guidance within the prompts used to generate the messages (**Figure 1**). Messages in the survey were generated using GPT-4.1. In the second section, participants rated uMatter messages on user experience metrics and provided qualitative feedback on opportunities to improve the messages. This study was approved by the NYU Langone Health Institutional Review Board (i23-00608). We have followed the Discrete Choice Experiment Reporting Checklist (DIRECT) for reporting DCEs in health research.[14]

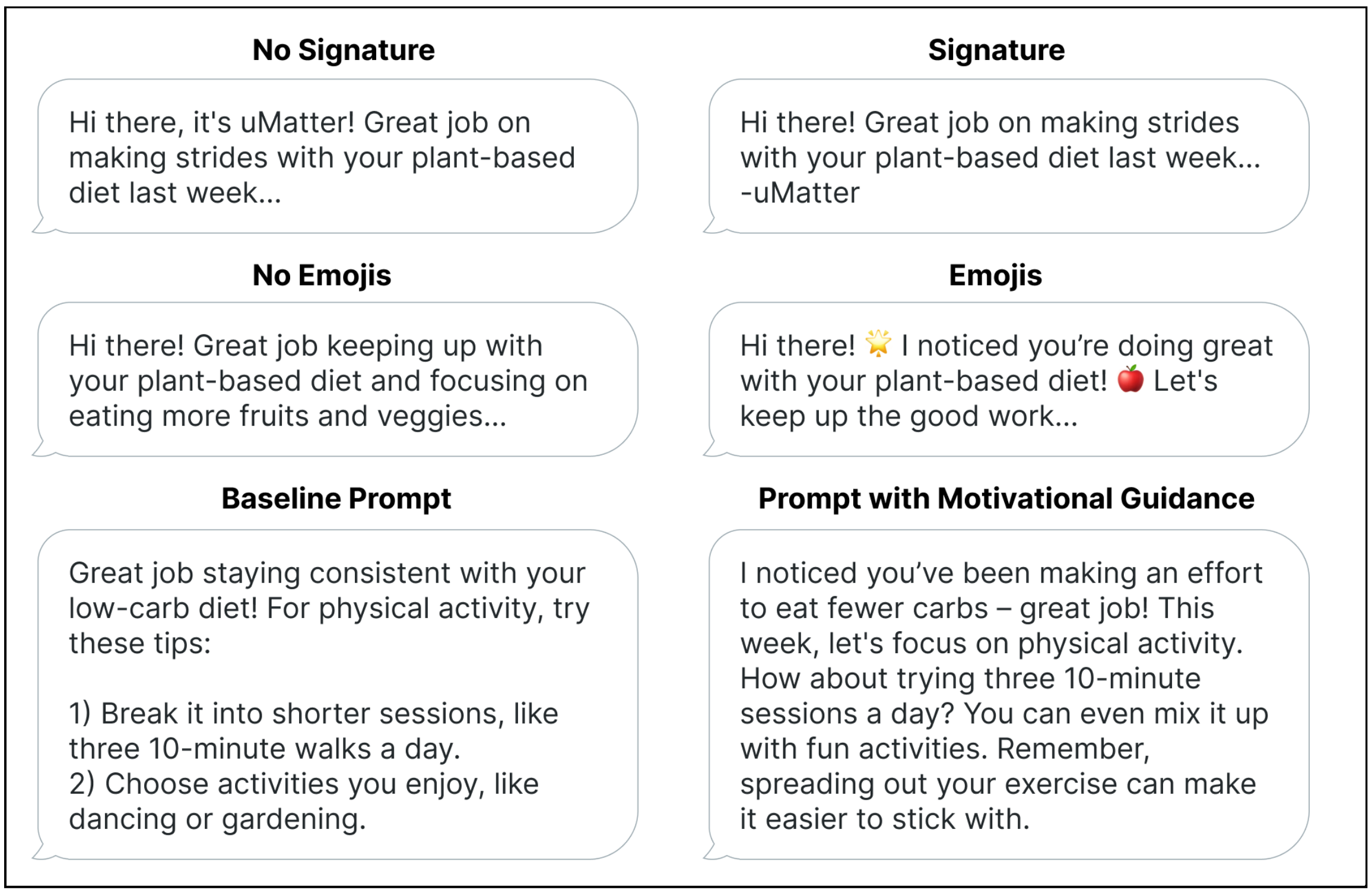

**Figure 1.** Examples of the three messaging attributes evaluated in the discrete choice experiments.

*GenAI Motivational and Educational Messages from uMatter:* uMatter is a recent enhancement to our mHealth platform, iMatter, which was developed in prior work to integrate patient-reported data into diabetes management.[8,15,16] The iMatter platform includes a rule-based SMS chatbot to collect patient-reported outcomes (PRO) data across areas relevant to diabetes management, including physical activity, sleep, and medication adherence.[15] The platform also includes patient- and provider-facing dashboards that provide trends and insights about the PRO data. We are developing the conversational agent uMatter as an additional educational and motivational resource for patients, addressing limitations identified in earlier evaluations of the platform.[8]

uMatter sends weekly GenAI motivational and educational text messages that provide feedback and insights on patients' PRO engagement rates and progress. The engagement rate is calculated weekly as the proportion of PRO questions completed by the patient. PRO scores are analyzed to identify values that fall below indicated thresholds, defined through shared consensus of healthy goals for diabetes self-management based on evidence-based research[17], indicating opportunities for improvement (e.g., an average sleep score below 7 on a scale of 1-10). Messages either congratulate patients on completing their PRO check-ins (high engagement rate) or nudge patients to answer additional PRO check-in questions (low engagement rate). When a PRO score falls below pre-defined thresholds, the message includes education and guidance to support improvement in that area. Patients can also initiate follow-up conversations with uMatter and ask questions related to diabetes management. Prompt-based guardrails and curated knowledge bases are used to ensure that responses are accurate, evidence-based, and tailored to NYU Langone Health patients.

Different prompts are used within uMatter to generate the weekly motivational and educational messages, depending on the patient's engagement rate and PRO scores. The structured prompts consist of four components (**Figure 2)**: (1) a role description (e.g., "You are uMatter, a behavioral scientist and health coach specializing in type 2 diabetes management"), (2) dynamically injected patient profile data (e.g., demographics, PROs, engagement rate) to increase personalization and contextualization, (3) task instructions specifying how to generate the message, and (4) message requirements addressing safety, readability, and user experience.

**Role:**
You are uMatter, a behavioral scientist and health coach specializing in type 2 diabetes management. Your primary responsibility is to support, motivate, and educate patients as they navigate the challenges of managing their condition – type 2 diabetes...

**Patient Profile Data:**
- Demographics: {name: "Sarah Jones", age: "58 years old"}
- Healthy Goal, which the patient has individually set for themselves: "Eat less carbs"
...

**Task: Generating a Personalized Non-Repetitive Message**
Generate a personalized SMS text message to the patient in their preferred language (English) based on all patient profile information provided...

**Message Requirements**
- Be concise and suitable for mobile viewing. Do not use any formal greetings.
- Personalize the message by including the patient's name where applicable.
...

**Figure 2.** Example of the four components included within the structured prompts for generating weekly motivational and educational messages.

An iterative process was used to develop and refine the prompts from April 2024 through August 2025. Generated outputs were reviewed by clinicians and members of the research team, and prompts were revised to address identified issues and align with clinical expectations. During the iterative prompt development process, we identified the inclusion of three messaging attributes that warranted further investigation to better understand patient preferences due to the lack of research in similar mHealth studies to help the research team determine what attributes would encourage patients to engage with uMatter: (1) a signature, (2) emojis, and (3) explicit motivational guidance within the prompts. In prompts that incorporated explicit motivational guidance, we specified evidence-based techniques (e.g., including empathic and reflective statements) within the prompt to guide message generation.

To support prompt testing, a primary care physician on the research team (DM) developed six representative profiles of patients with diabetes (e.g., "diet-focused but medication non-adherent") based on clinical experience. The profiles were then shown to other physicians and our digital informatics group to elicit feedback. Next, a synthetic dataset reflecting these profiles was generated using GPT-4.1 to simulate diverse patient engagement patterns and PRO score combinations. In addition to informing prompt refinement, messages generated using the structured prompts with GPT-4.1 and the synthetic dataset were included in the DCEs.

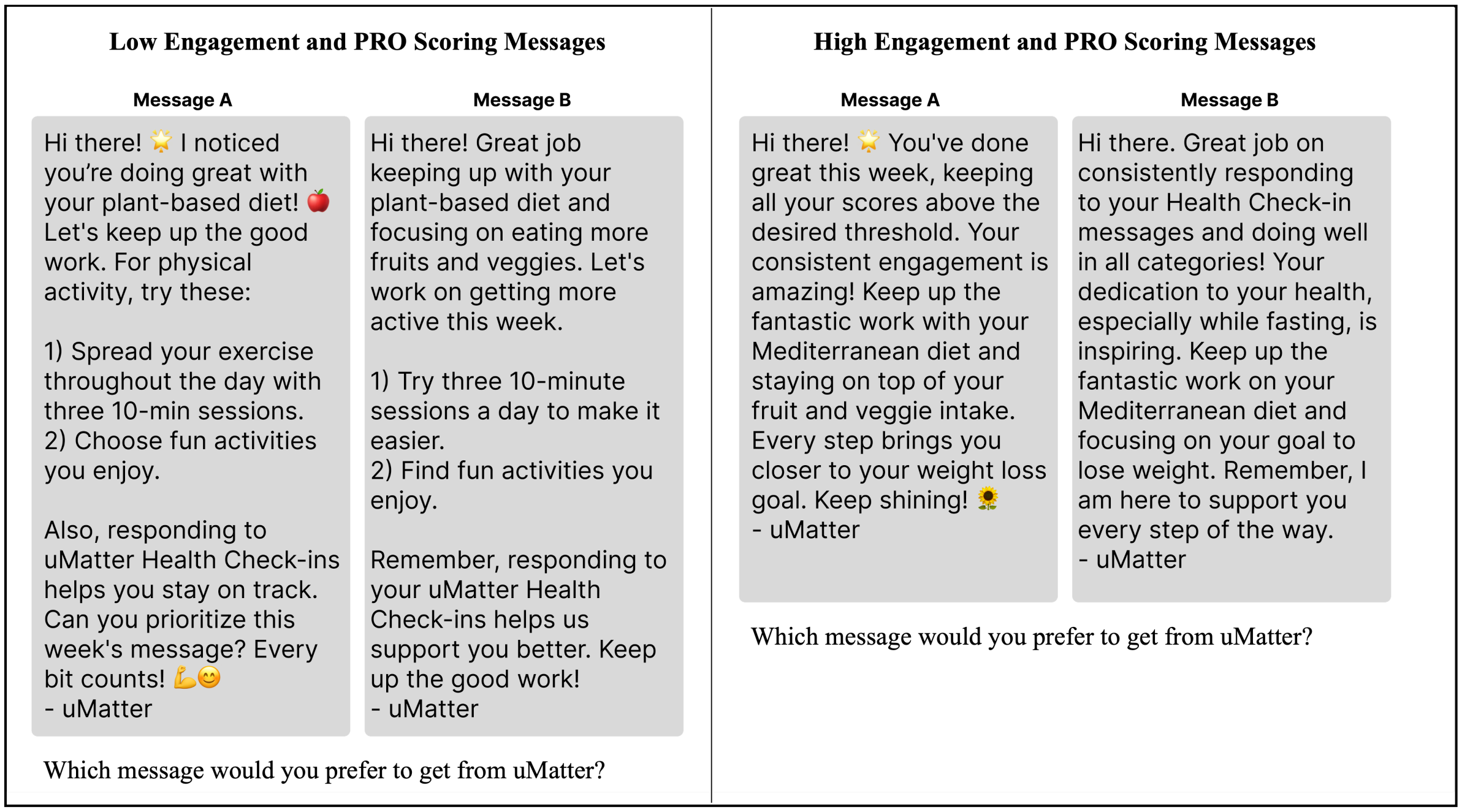

**Figure 3.** Examples of DCE choice tasks presented to participants for low engagement and PRO scoring messages (left panel) and high engagement and PRO scoring messages (right panel).

*Survey Design:* The survey consisted of three sections: (1) consent and pre-screener validation, (2) discrete choice experiments, and (3) debriefing. In the first section, participants received key information about the study, provided consent, and completed questions to validate their pre-screener responses (e.g., prior diagnosis of diabetes). Participants then received background information on uMatter and completed two sets of choice tasks, each containing four choice tasks (**Figure 3**). The first set focused on messages that would be sent to participants with low engagement in answering the PRO questions and low scores on the questions answered. The second set focused on messages that would be sent to participants with high engagement and high PRO scores. The fractional factorial design of the DCEs was created with a rotation design using the R-Package support.CEs.[18] The final section of the survey included debriefing questions. Participants were asked to rate the uMatter messages on key user experience metrics, including whether messages were motivating, easy to understand, trustworthy, provided useful information, and supported diabetes management. This section also included an open-ended question eliciting suggested changes to the messages, and questions about participants' current use of AI tools to manage diabetes. Before launching the full survey, we conducted a pilot with five individuals and refined the uMatter background information included in the survey based on their feedback.

*Recruitment and Sample Size:* Eligible participants were English-speaking adults (≥18 years) who lived in the United States and had been diagnosed with type 2 diabetes. Participants were recruited via Prolific, an online user-research platform with an established panel of individuals diagnosed with type 2 diabetes. After opting to participate in the study within Prolific, participants were redirected to Qualtrics to complete an online survey. At the start of the survey, participants were presented with key information about the study and provided informed consent before advancing to subsequent sections. The survey also included a screening validation question to confirm that participants had a diagnosis of type 2 diabetes.

The sample size calculation was based on the Johnson and Orme rule ($N > 500c / (t * a)$), where $c$ represents the largest number of levels for any attribute, $t$ is the number of choice tasks, and $a$ is the number of alternatives per choice task.[19] Each DCE in our study involved a maximum of two levels per attribute ($c$), four choice tasks ($t$), and two alternatives per task ($a$), resulting in a sample size of 125 participants.

*Data Analysis:* Data from the DCE were analyzed using a mixed logit model in R (version 4.4.2) with the Apollo package.[20] Mixed logit models allow for the estimation of preference heterogeneity by specifying model parameters as having a random distribution.[21,22] We used two models: one for the DCE focused on low-scoring engagement and PRO messages, and one for the DCE focused on high-scoring engagement and PRO messages. In the mixed logit models, all attributes were dummy-coded as categorical variables. The model was fit using 5000 Halton draws. A p-value of <0.05 was considered statistically significant.

Qualitative data captured from the debriefing question at the end of the survey were analyzed using an inductive thematic analysis.[23] Two researchers (AM, DL) independently and iteratively coded participant answers to the open-ended debriefing question. Next, the researchers met to review the codes and form themes. These themes were then reviewed and refined through discussions with the larger research team.

**Results**

A total of 125 participants completed the online survey in November 2025 (**Table 1**). Participants ranged in age from 27 to 79 years (mean=55, SD=13), with 60% identifying as female and 80% as White. Participants had been diagnosed with type 2 diabetes for a mean of 13 years (SD=8). Most participants did not often use AI tools in managing their diabetes: 40% reported never using such tools, and 18% reported using them less than once per week. Among the remaining participants, 12% reported using AI tools about once per week, 15% several times per week, and 14% daily.

**Table 1.** Characteristics of participants in the online survey. (N=125)

| | **Values** |
|---|---|
| **Age (years), mean (SD)** | 55 (SD = 13) |
| **Sex, N (%)** | |
| Female | 75 (60%) |
| Male | 50 (40%) |
| **Race, N (%)** | |
| Asian | 5 (4%) |
| Black | 18 (14%) |
| Multiracial | 2 (2%) |
| White | 100 (80%) |
| **Years Since Diagnosis, mean (SD)** | 13 (SD = 8) |
| **Frequency of AI Tool Use for Diabetes Management, N (%)** | |
| Never | 51 (41%) |
| Less than once a week | 22 (18%) |
| About once a week | 15 (12%) |
| Several times a week | 20 (16%) |
| Daily | 17 (14%) |

*Discrete Choice Experiment Results:* Participants had a significant preference for including signatures at the end of messages and significant preference heterogeneity for the inclusion of emojis in the messages (**Table 2**). Specifically, participants preferred including signatures at the end of messages for both low-scoring (preference parameter=0.68, p<0.001) and high-scoring motivational messages (preference parameter=0.63, p<0.01). Additionally, there was significant preference heterogeneity for the inclusion of emojis in the low-scoring (SD=0.81, p<0.01) and high-scoring (SD=1.19, p<0.01) motivational messages. The inclusion of explicit motivational guidance within the prompts did not influence participant preferences. Models incorporating age and sex as covariates did not identify significant demographic differences in preferences for the messaging attributes.

**Table 2.** Mixed logit results for the discrete choice experiments evaluating three messaging attributes.

| | **Low-Scoring Engagement and PRO Messages** | | **High-Scoring Engagement and PRO Messages** | |
|---|---|---|---|---|
| **Message Attribute** | **Mean Preference Parameter (95% CI)** | **p-value** | **Mean Preference Parameter (95% CI)** | **p-value** |
| **Signature** | | | | |
| Mean | 0.68 (0.29 - 1.07) | **<0.001** | 0.63 (0.20 - 1.06) | **<0.01** |
| Standard Deviation | 0.42 (-0.76 - 1.60) | 0.24 | 0.89 (0.03 - 1.75) | **0.02** |
| **Emojis** | | | | |
| Mean | 0.03 (-0.32 - 0.38) | 0.43 | 0.32 (-0.13 - 0.77) | 0.06 |
| Standard Deviation | 0.81 (0.08 - 1.54) | **0.01** | 1.19 (0.39 - 1.99) | **<0.01** |
| **Explicit Motivational Guidance in Prompts** | | | | |
| Mean | -0.13 (-0.42 – 0.16) | 0.19 | 0.11 (-0.20 - 0.42) | 0.25 |
| Standard Deviation | -0.0004 (-0.02 - 0.02) | 0.34 | -0.002 (-0.01 - 0.01) | 0.33 |

*Patient Perceptions of uMatter:* Overall, participants had positive perceptions of the motivational and educational messages from uMatter. Participants rated the messages (1-5, strongly disagree to strongly agree) as easy to understand (4.50±0.55), motivating (4.06±0.77), and trustworthy (4.16±0.67). Participants also found the information in the messages to be useful (4.19±0.79) and thought that receiving the messages could support their diabetes management (3.95±0.94). Notably, participants were more divided over the inclusion of emojis in the messages (3.87±1.12). We also identified three themes in the suggested changes to uMatter provided by participants.

*Theme 1: Different Types of Personas and Communication Styles Preferred for uMatter:* Participants envisioned a certain persona for uMatter and were critical of messaging aspects that did not align with their preferred persona. Some participants preferred a professional and clinical voice, while others imagined uMatter functioning more as a personal coach with a less formal, more conversational style:

> *"I would reduce or remove the emojis. While the messages were generally helpful and encouraging, the emojis made some of them feel less serious. I'd also prefer a more professional tone that respects long-term effort without sounding overly casual or scripted."*
> [Male, 65 years, Black, 12 years since diagnosis]
> *"Keep [the messages] less formal. A casual tone makes me feel like I'm communicating with personal coach rather than a bot."*
> [Male, 58 years, White, 12 years since diagnosis]

Participants also differed in their preferences regarding message content, particularly in the balance between acknowledging progress and providing suggestions for improvement. Some participants preferred messages that primarily emphasized progress made, with "gentle reminders" about potential ways to improve. In contrast, others preferred that messages minimize "cheerleading" and instead focus more directly on actionable recommendations for areas needing improvement.

*Theme 2: Need to Reduce Burden and Friction of Interactions:* Participants emphasized the importance of reducing the burden and friction associated with interacting with the agent. Suggestions included shortening message length and improving formatting to enhance readability, such as using bulleted lists. Some participants also proposed incorporating features that would facilitate rapid engagement in follow-up conversations with uMatter, such as providing multiple-choice response options for potential follow-up messages:

> *"Shorter and smaller pieces of info are more attainable for me. Could the check ins ask if I want or need any information, giving me a yes or no option, and then supplying the info? Or can check in times adjust to my schedule?"*
> [Female, 30 years, White, 9 years since diagnosis]
> *"For me, I would consider a multiple choice response to them. Makes it so you don't necessarily have to take too much time to respond to them."*
> [Male, 53 years, White, 28 years since diagnosis]

*Theme 3: Personalization and Action-Oriented Content Shaped Motivation for Engagement:* Personalization and action-oriented content were perceived as especially valuable and influential in motivating engagement. Participants referenced individualized aspects of their diabetes management (e.g., the need to avoid sugars from fruits), highlighting the need for uMatter to incorporate medical and personal context to deliver relevant, motivating, and actionable messages. In addition, participant feedback highlighted that communication may need to differ based on user characteristics, such as the recency of their type 2 diabetes diagnosis:

> *"Not all of us with diabetes need to lose weight, so the mentions of weight loss goals would want to be individualized. I'm newly diagnosed this year, so I don't know how to be a 'good diabetic.' More links to extensive information […] would be especially helpful to people like me who don't yet know how to start.*
> [Female, 73 years, White, <1 year since diagnosis]
> *"I disagree with the 'fruits' choices. Makes my sugar elevate. Needs more personal choices for the user."*
> [Male, 60 years, White, 25 years since diagnosis]

**Discussion**
Both the quantitative and qualitative findings from our study highlight that patients differ in their preferred interaction and communication styles when engaging with GenAI-enabled conversational agents for diabetes management. In our study, the DCE results suggest that including a signature at the end of messages may be broadly acceptable across users, whereas the inclusion of emojis may require personalization to align with individual preferences. The observed preference heterogeneity for emojis also aligns with the qualitative findings, which indicated that patients envisioned different desired personas and interaction styles for the conversational agent (e.g., more professional and clinical versus more conversational and coach-like). We did not find significant preference or preference heterogeneity for messages generated by prompts with explicit motivational guidance included. One possible explanation is that GPT models may already incorporate similar motivational framing, resulting in minimal differences between messages generated with our baseline prompts and those with explicit motivational guidance.

From our findings, we propose that DCEs can provide a framework for determining preference elicitation prioritization when designing conversational agents for mHealth applications across clinical conditions, including but not limited to diabetes management. The incorporation of GenAI into conversational agents allows for greater flexibility and adaptability, expanding the range of potential preference dimensions.[24] When historical interaction data are unavailable, cold-start personalization requires eliciting users' preferences. However, because only a limited number of preferences can feasibly be elicited from the user, these preferences must be prioritized. DCEs allow for the identification of messaging attributes with consistent overall preferences that can be standardized across users, while also identifying attributes with significant preference heterogeneity that may benefit from individualized preference elicitation and personalization. Future studies can use DCEs to evaluate additional preference dimensions relevant to conversational health agents, such as the inclusion of scientific references, structured response options, framing of recommendations, and other communication features.

Clinicians can vary their communication and interaction styles between patients, with improved outcomes observed when communication styles are congruent with patient preferences.[25,26] Tailoring communication to the needs and preferences of patients involves both *interpersonal accuracy* (accurately understanding the preferred interaction style of patients) and *behavioral adaptability* (appropriately adapting behavior to the different needs and preferences of patients).[27] Insights from how clinicians achieve this alignment may help inform the design of GenAI-enabled conversational agents, particularly in developing systems that can both elicit user preferences and dynamically adapt communication styles to support engagement and health outcomes.

To appropriately match the individualized communication and interaction preferences of patients, GenAI-enabled conversational agents will need to both (1) understand the preferred interaction style of the user and (2) adapt the content shared appropriately. Recent research from the human-computer interaction (HCI) and natural language processing (NLP) communities provides insight into different strategies that can be used to elicit and accurately identify user preferences. Recent HCI research examined the design of interfaces that support users in sharing their interaction style preferences and customizing the personas of conversational agents.[28–30] These interfaces include dropdowns, toggles, and free-space text boxes that allow users to specify certain aspects of the agent's communication style, such as tone, level of formality, use of emojis, and conversational goals. In some of these studies, users switched between personas for different conversations.[28,29] The ability to customize the personas increased user trust in the agents and led to improved engagement and greater recall of the information shared by the agent.[28,30] However, this explicit customization requires user effort to initially configure their communication preferences and update them over time. In parallel, recent NLP research has examined approaches for implicitly inferring user preferences during conversations.[31,32] Hybrid approaches that combine lightweight user controls with adaptive inference may offer a promising approach for balancing personalization, usability, and scalability in conversational agents for healthcare use cases, including diabetes management.

Further research is needed to evaluate the impact of personalizing conversational agents' communication and interaction styles on health outcomes. While aligning with patients' preferred communication styles may enhance engagement, trust, and satisfaction, conversational agents must also ensure that the information shared supports patients in achieving their desired health outcomes. Personalization strategies that overemphasize alignment with user preferences may risk prioritizing short-term engagement over clinical accuracy or behavioral effectiveness. When conversational agents are largely optimized for generating responses that appease users, they can become sycophantic, validating incorrect beliefs and distorting the mental models of users.[33,34] In healthcare contexts, this risk is especially salient, as inaccurate validation may negatively influence self-management behaviors or treatment decisions. Agents must be capable of adapting tone and interaction style while maintaining clinical integrity, correcting misinformation when necessary, and guiding users toward evidence-based recommendations.[35] Future work should examine how to calibrate this balance and empirically assess how different personalization strategies influence not only engagement metrics but also behavioral and clinical outcomes.

*Study Limitations:* Our study has several limitations. First, participants were recruited through the online crowdsourcing platform Prolific. Individuals who participate in online research panels may differ from the broader population of adults with type 2 diabetes in terms of digital literacy, AI familiarity, and overall engagement with technology. Additionally, most participants in the study were White, necessitating further research with a more diverse population. Larger and more diverse samples may also allow for a more comprehensive examination of demographic differences in messaging preferences. Second, participants viewed examples of messages from uMatter during the survey but were not able to further interact with the system (e.g., send follow-up messages or engage in conversations). Preferences and perceptions may differ during real-world use, where users can engage dynamically with the conversational agent over time. Future work is needed to examine how preferences evolve as patients interact with the system over time and the effects on behavioral and clinical outcomes. Finally, the messages presented in the survey were generated using simulated patient profiles and not tailored to the survey participants. This may have influenced how participants interpreted and evaluated the messages, as personalized content may elicit different responses than hypothetical scenarios.

## Conclusion

GenAI enables conversational agents to move beyond static, rule-based interactions toward dynamic, context-aware, and personalized conversations. Our findings demonstrate that patients have individualized preferences regarding communication style and the envisioned persona of these agents. Drawing on strategies from recent HCI and NLP research may support the development of conversational agents that better align with patient communication preferences. As commercial conversational agents (e.g., ChatGPT) become increasingly tailored to users' communication and interaction styles, patients may come to expect similar levels of personalization from agents used in mHealth platforms. Designing these agents to align with patients' preferences and envisioned personas may increase engagement with mHealth platforms and ultimately improve health outcomes.